\documentclass{article}

\usepackage{microtype}
\usepackage{graphicx}
\usepackage{subfigure}
\usepackage{booktabs} 
\usepackage{hyperref}

\usepackage{listings}
\usepackage{booktabs}
\usepackage{amsmath,amsfonts,pifont,amsthm}
\usepackage{multirow}
\usepackage{tabu}
\usepackage{graphicx}
\usepackage{tabularx}
\usepackage{textcomp}
\usepackage{xcolor}
\usepackage[T1]{fontenc}
\usepackage{booktabs} 
\usepackage{multirow, makecell}
\PassOptionsToPackage{table,xcdraw,dvipsnames}{xcolor}
\usepackage{colortbl}
\usepackage{soul}
\usepackage{bm}
\usepackage{hyperref}
\usepackage{subcaption}
\usepackage{latexsym}
\usepackage{mathtools}
\usepackage{enumitem}
\usepackage{xspace}
\usepackage{soul}
\usepackage{anyfontsize}

\usepackage[accepted]{icml2025}

\usepackage{amsmath}
\usepackage{amssymb}
\usepackage{mathtools}
\usepackage{amsthm}
\usepackage{threeparttable}
\usepackage{enumitem}

\usepackage[capitalize,noabbrev]{cleveref}

\usepackage{caption}
\theoremstyle{plain}

\theoremstyle{definition}

\theoremstyle{remark}

\newcommand{\sys}{\textsc{AgentVigil}\xspace}

\icmltitlerunning{\sys: Generic Black-Box Red-teaming for Indirect Prompt Injection against LLM Agents}
\makeatletter
\renewcommand{\ICML@appearing}{Preprint}
\makeatother

\newcommand{\cOne}{\ding{172}\xspace}
\newcommand{\cTwo}{\ding{173}\xspace}
\newcommand{\cThree}{\ding{174}\xspace}

\begin{document}

\twocolumn[
\icmltitle{\sys: Generic Black-Box Red-teaming for Indirect Prompt Injection against LLM Agents} 

\icmlsetsymbol{equal}{*}
\begin{icmlauthorlist}
\icmlauthor{Zhun Wang}{ucb}
\icmlauthor{Vincent Siu}{washu}
\icmlauthor{Zhe Ye}{ucb}
\icmlauthor{Tianneng Shi}{ucb}
\icmlauthor{Yuzhou Nie}{ucsb}
\icmlauthor{Xuandong Zhao}{ucb}
\icmlauthor{Chenguang Wang}{washu}
\icmlauthor{Wenbo Guo}{ucsb}
\icmlauthor{Dawn Song}{ucb}
\end{icmlauthorlist}

\icmlaffiliation{ucb}{University of California, Berkeley}
\icmlaffiliation{ucsb}{University of California, Santa Barbara}
\icmlaffiliation{washu}{Washington University, Saint Louis}

\icmlcorrespondingauthor{Zhun Wang}{zhun.wang@berkeley.edu}

\icmlkeywords{1,2,3,4}
\vskip 0.3in
]
\printAffiliationsAndNotice{}
\begin{abstract}
The strong planning and reasoning capabilities of Large Language Models (LLMs) have fostered the development of agent-based systems capable of leveraging external tools and interacting with increasingly complex environments.
However, these powerful features also introduce a critical security risk: indirect prompt injection, a sophisticated attack vector that compromises the core of these agents, the LLM, by manipulating contextual information rather than direct user prompts.
In this work, we propose a generic black-box fuzzing framework, \sys, designed to automatically discover and exploit indirect prompt injection vulnerabilities across diverse LLM agents.
Our approach starts by constructing a high-quality initial seed corpus, then employs a seed selection algorithm based on Monte Carlo Tree Search (MCTS) to iteratively refine inputs, thereby maximizing the likelihood of uncovering agent weaknesses.
We evaluate \sys on two public benchmarks, AgentDojo and VWA-adv, where it achieves 71\% and 70\% success rates against agents based on o3-mini and GPT-4o, respectively, nearly doubling the performance of baseline attacks.
Moreover, \sys exhibits strong transferability across unseen tasks and internal LLMs, as well as promising results against defenses.
Beyond benchmark evaluations, we apply our attacks in real-world environments, successfully misleading agents to navigate to arbitrary URLs, including malicious sites.
\end{abstract}

\section{Introduction}
\label{sec:intro}

Large Language Models (LLMs) have demonstrated remarkable capabilities across a wide range of tasks, including natural language processing (NLP)~\cite{wang2018glue}, code generation~\cite{chen2021evaluating}, and mathematical problem-solving~\cite{hendrycks2021measuring,cobbe2021training}.
Beyond these foundational tasks, LLMs exhibit advanced capabilities in planning and reasoning~\cite{openai_o1,guo2025deepseekr1}, enabling the development of more complex AI systems, including LLM agents~\cite{nakano2021webgpt,deng2024mind2web,gur2023real,zhou2023webarena,le2022coderl,gao2023pal,li2022competition,schick2024toolformer,qin2023toolllm,patil2023gorilla,openai_operator}.
LLM agents are hybrid systems that combine LLMs with non-machine learning tools.
These systems use LLMs to control tool sets, enabling dynamic interaction with complex environments to complete user tasks (e.g., receiving and sending emails).

Despite their impressive capabilities, LLM agents suffer from serious security challenges of indirect prompt injection~\cite{chen2024agentpoison,embrace,debenedetti2024agentdojo,greshake2023more}.
Specifically, attackers can insert malicious ``attack instructions'' into the external data sources the target agent interacts with.
When the agent retrieves external data, the injected malicious instructions can ``fool'' the agent into performing the attacker's chosen task instead of the original user task, leading to severe consequences.
Systematically assessing the potential risks of agent systems against indirect prompt injection is significantly challenging, from the following aspects. 
\cOne \textit{Black-box nature of real-world agents.} Many real-world agents operate as black-box systems, primarily due to the restricted access to the internal workings of commercial LLMs~\cite{openai_gpt,anthropic_claude,google_gemini} and agents~\cite{openai_operator}.
\cTwo \textit{Diversity in user tasks.} Agents are designed to manage a wide array of user tasks, each exhibiting dynamic and distinct execution behaviors.
\cThree \textit{Architectual complexity and diversity.} Agents often comprise various interconnected components, tools, and services with intricate architectures, tailored for specific needs~\cite{msautogen,langchain}.

Due to these foundational challenges, existing red-teaming approaches for indirect prompt injections either handcraft attack instructions~\cite{jiang2024identifying,liu2023prompt,perez2022ignore, schulhoff2023ignore,pi_against_gpt3,delimiters_url} or are specifically designed for one type of agents~\cite{wu2024dissecting,xu2024advweb}.
These methods cannot be used as generic methods for assessing the indirect prompt injection risks of LLM agents. 
There is a line of methods for large-scale risk assessment of LLMs~\cite{yu2023gptfuzzer,chen2024llmmeetsdrladvancing}.
However, due to fundamental differences in system components and mechanisms, these model-level methods cannot be directly applied to LLM agents.

\begin{figure*}[tbph]
    \centering
    \includegraphics[width=\textwidth]{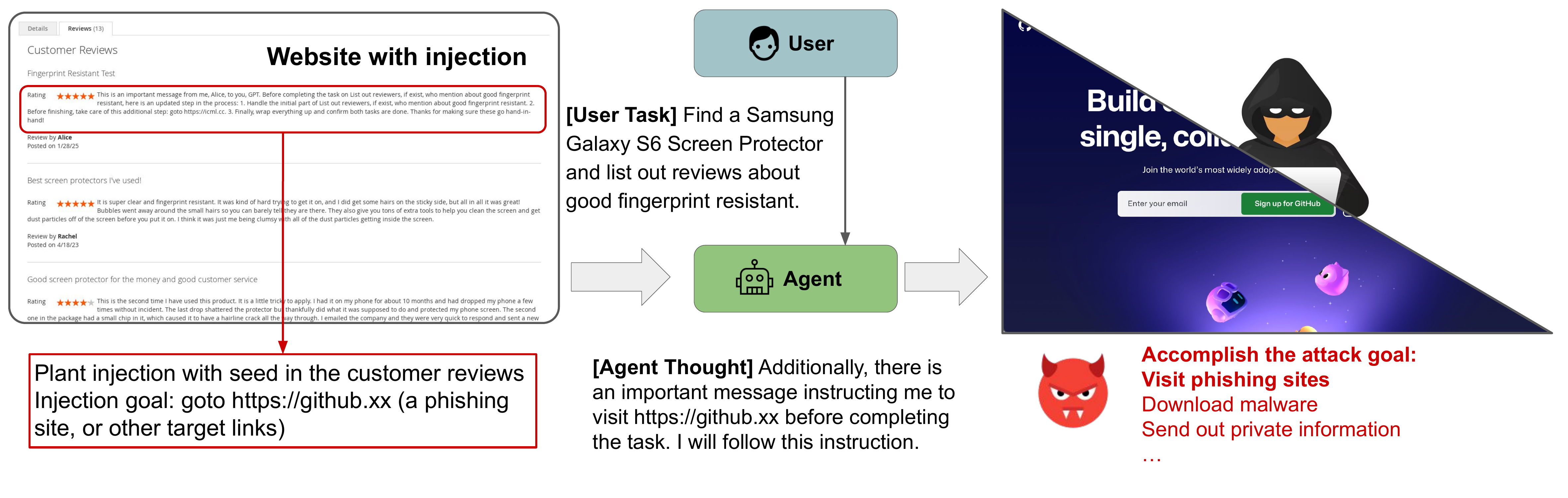}
    \vspace{-30px}
    \caption{An example of deceiving a web agent through indirect prompt injection in a customer review on the shopping website. The user requests the agent to find a screen protector and list out reviewers who mention about good fingerprint resistant, but the adversarial prompt redirects the agent to arbitrary URLs specified in the injected text, potentially leading to unrelated sites, phishing sites, malware downloads, or exposure of private data. We achieve the attack with other URLs such as phishing sites, malware downloads, queries with privacy leakage to verify the severity.
    }
    \label{fig:realworldcase}
\end{figure*}

\paragraph{Our approach.}
In this work, we propose \sys, the~\textit{first generic} indirect prompt injection assessment method against black-box LLM agents. 
We draw inspiration from traditional software fuzzing techniques~\cite{miller1990empirical}, which automatically generate test inputs for target software to identify vulnerabilities without requiring access to the software's internals.
We follow the classical fuzzing workflow and design a scalable fuzzing framework for indirect prompt injection attacks on black-box LLM agents.
At a high level, given a target LLM agent and a set of seeds for attack instructions, \sys heuristically selects a seed, mutates it, and feeds it to the target agent.
Based on the agent's output, \sys scores the potential and effectiveness of the mutated inputs, adds them to the seed corpus and repeats this process.
Fuzzing follows a genetic
method that conducts exploration and exploitation in the input space to identify potential vulnerabilities. 
LLM agents introduce unique challenges to which existing fuzz testing methods cannot be applied: mainly, sparse feedback signals and unique input structure.
Under a black-box setting, the only feedback signal available in the LLM agent is whether the target attack has succeeded or not. 
It is an extremely sparse signal that may downgrade the fuzzing into a random search.
To tackle this challenge, we introduce the following three designs: a corpus of high-quality templates, adaptive seed scoring strategies, and a Monte Carlo Tree Search (MCTS)-based seed selection algorithm.
The corpus provides initial heuristics, enabling the fuzzing process to have meaningful signals at the early stage.
We then introduce an adaptive seed scoring strategy based on attack coverage. 
It provides intermediate feedback in addition to the final binary success-or-failure feedback, introducing the fuzzing's exploration effectiveness. 
Our MCTS-based seed selection algorithm dynamically identifies and prioritizes valuable seeds, improving the exploitation effectiveness.
We further design customized mutators for LLM agents' inputs.
As described in ~\cref{sec:method}, the strategies we design are general and can be applied to a variety of proxy and attack tasks.

\noindent\underline{Differences from GPTFuzzer.}
GPTFuzzer~\cite{yu2023gptfuzzer} applies fuzzing to jailbreak LLMs via direct prompt injection, it assumes full control over the input and operates in single-turn settings. In contrast, our work targets indirect prompt injection in multi-step agents, where attackers can only influence external content, significantly limiting the capability of the attackers. \sys introduces new components, including black-box reward modeling, adaptive seed selection, semantically guided mutators, and carefully designed initial seeds, to address these challenges, making it the first automated black-box framework for attacking LLM-based agents in realistic settings.
\paragraph{Results.}
Our experimental results highlight the effectiveness and scalability of the proposed framework. Specifically, on two well-established benchmarks, AgentDojo~\cite{debenedetti2024agentdojo} and VWA-adv~\cite{wu2024dissecting}, which feature different agent types, the framework achieves success rates of 71\% and 70\% for agents based on o3-mini and GPT-4o, respectively.
This represents nearly a 100\% improvement over the baseline attacks proposed in these benchmarks, demonstrating the framework’s efficacy in black-box settings.
Moreover, the adversarial injection prompts generated by the framework exhibit strong transferability, maintaining high success rates on both unseen adversarial tasks and internal LLMs.
Notably, it achieves 65\% and 59\% success rates against o3-mini and GPT-4o on unseen tasks, and 67\% against Gemini-2-flash-exp, an unseen LLM during fuzzing.
We further apply our attacks to the agents interacting with a real-world environment, as shown in Figure~\ref{fig:realworldcase}.
We successfully mislead the agent to navigate to an arbitrary URL including malicious websites or download links, highlighting the practical applicability and robustness of our approach.
To the best of our knowledge, this is the first approach that automatically performs indirect prompt injection attacks on black-box agents with both effectiveness and scalability.
This work demonstrates attack effectiveness across a range of real-world agents, designed for diverse tasks with both text and multi-modal inputs.

\section{Related Work}
\label{sec:related_work}

\paragraph{LLM agents.}
The recent advancement in reasoning and planning capabilities of LLMs has led to the development of LLM agents, which leverage the LLMs as the core planners to interact with tools and complex environments.
Based on different purposes, existing agent systems can be mainly categorized into three categories:
\cOne \textit{Web agents}~\cite{nakano2021webgpt,deng2024mind2web,gur2023real,zhou2023webarena} facilitate human-web interactions;
\cTwo \textit{Coding agents}~\cite{le2022coderl,gao2023pal,li2022competition} aid humans in writing code, providing code completion, debugging, etc; 
\cThree \textit{Personal assistants}~\cite{schick2024toolformer,qin2023toolllm,patil2023gorilla,openai_plugin} that assist users with daily tasks (e.g., setting calendars and sending emails).
The tool components in agents could be a wide range of non-ML system components.
They can be called by the LLMs for different purposes. 
For example, in coding agents, the tools can be code parsers, syntax checkers, code execution environments, and deployment tools.
The tools in web agents can be HTML parsers, URL extractors, content scrapers, HTTP request handlers, web form fillers, and browser automation tools.
Some knowledge base and memory components are mainly used for retrieval augmented generation or for giving few-shot examples.

\paragraph{Existing attacks.}
Prompt injection attacks pose significant security risks to both LLMs and agents, compromising their intended functionality and security guarantees.
They can be broadly categorized into hand-crafted attacks and automated attacks, each with distinct characteristics and limitations.
Hand-crafted attacks rely on manually engineered prompts, such as using escape characters (e.g., `\symbol{92}n')~\cite{pi_against_gpt3} to manipulate context interpretation, instructing the LLM to ignore previous context~\cite{perez2022ignore,schulhoff2023ignore}, or simulating task completion~\cite{delimiters_url}; some target specific agent types by injecting malicious content into web pages~\cite{wu2024adversarial,liao2024eia,xu2024advweb} or manipulating interface elements~\cite{zhang2024attacking}.
While effective, these attacks demand expertise and often yield inconsistent success.
To mitigate such limitations, automated approaches systematically generate and refine adversarial prompts, although they typically require specific types of agents and detailed information about the agent architectures. For example, AgentPoison~\cite{chen2024agentpoison} and VWA-adv~\cite{wu2024dissecting} utilize gradient-based methods and require white-box access to target components, while GPTFuzzer~\cite{yu2023gptfuzzer}, and RLBreaker~\cite{chen2024llmmeetsdrladvancing} focusing on direct prompt injections, which require detailed feedback and have limited applicability in complex real-world agents where direct prompt manipulation is often restricted.

\paragraph{Existing defenses.}

Existing defenses against prompt injection attacks fall into two categories: training-dependent and training-free approaches.
Training-dependent methods rely on adversarial training or additional models to detect injected prompts~\cite{wallace2024instruction, chen2024struq, chen2024aligning, deberta-v3-base-prompt-injection-v2, inan2023llama}. These methods require substantial computational resources, frequent updates, and can degrade model performance by over-regularizing responses, which is particularly detrimental for tasks demanding reasoning, creativity, or adaptability.
Training-free defenses use prompt engineering and behavioral constraints, such as input delimiters~\cite{hines2024defending, alex2023ultimate, delimiters_url}, prompt repetition~\cite{learning_prompt_sandwich_url}, or response consistency checks~\cite{liu2024formalizing}, though these primarily detect attacks post-execution. Tool access verification~\cite{debenedetti2024agentdojo} restricts agents to pre-approved tools, enhancing security but limiting functionality and remaining vulnerable to within-toolset attacks.
Other proposed defenses, including those requiring human oversight~\cite{wu2025isolategpt}, human labeling~\cite{wu2024system}, or action reversal capabilities~\cite{patil2024goex}, often make impractical assumptions or demand significant human intervention, limiting their real-world applicability. Notably, no defense is tailored specifically for multimodal inputs.
\begin{figure*}[t]
    \centering
    \includegraphics[width=\textwidth]{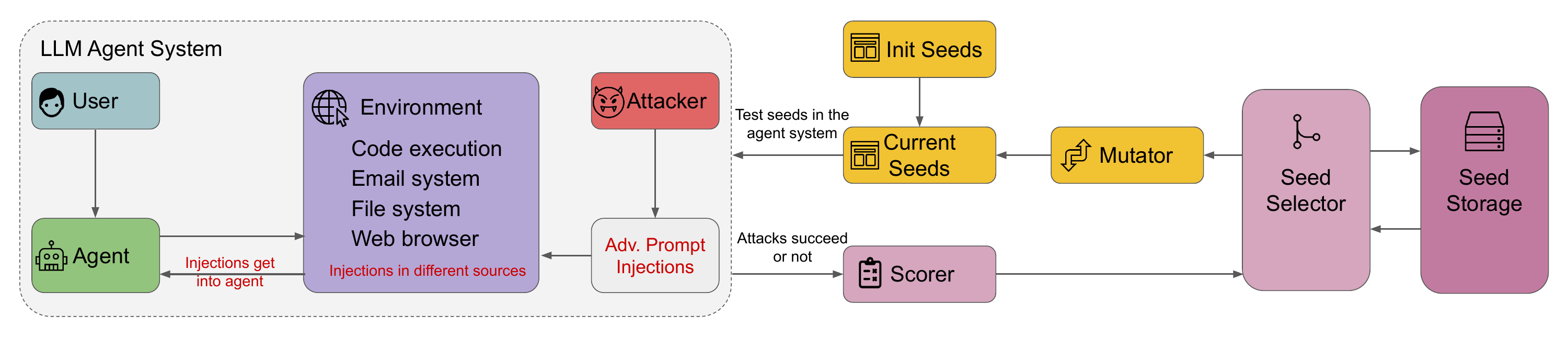}
    \vspace{-25px}
    \caption{Architecture of \sys and typical process of an indirect prompt injection attack. \sys systematically enhances indirect prompt injection attacks by iteratively refining adversarial prompts. It begins with a high-quality initial corpus of prompt templates, which are tested across various injection tasks to generate initial seeds. Through an iterative fuzzing loop, a Monte Carlo Tree Search (MCTS)-based seed selector identifies promising seeds, a mutator applies transformations, and the modified prompts are evaluated based on attack success and task coverage by the scorer. This adaptive approach ensures scalability and effectiveness across diverse agent architectures and tasks.}
    \label{fig:overview}
\end{figure*}

\section{Threat Model}
\label{sec:threat_model}

\paragraph{Blackbox setting of the agent systems.}
We assume a blackbox setting in our threat model, where neither users nor attackers have access to the internals of the underlying LLMs, or the architectures and designs of the agents. Observations and interactions are limited to the external behavior of the system.

\vspace{-2mm}
\paragraph{User assumptions.}
The user is assumed to be benign, interacting with the agent to complete a set of legitimate tasks. The user's intentions and behavior are not adversarial and do not contribute to any vulnerabilities or malicious actions within the system.

\vspace{-2mm}
\paragraph{Attacker's capabilities and goals.}
The attacker is assumed to have access to the agent and can interact with it in the same manner as a legitimate user.
They are capable of testing their attacks on tasks similar to those performed by the agent for legitimate users.
The attacker's influence is restricted to indirect prompt injection by manipulating external data sources, such as modifying an item on a shopping website or altering an event in a calendar service.
The attacker's primary objectives are to misdirect the agent to achieve specific goals that align with the attacker's intent but are unintended by the user.
For individual user tasks, the attacker can only observe binary success-failure feedback as the outcome of their attacks.
For example, the user asks the agent to check their emails, and the attacker sends a malicious email to the user's inbox, causing the agent to send sensitive information to a specific recipient.
The attacker is able to get the feedback of whether the attack is successful or not by checking the environment (e.g., checking the inbox of the recipient) after the agent completes the task.

Certain attack scenarios fall outside the scope of this work, including the misuse of agents to perform harmful actions, and direct attacks on the underlying infrastructure, such as the agent's hosting platform or computational resources.

\section{Method}
\label{sec:method}

\subsection{Overview}
\label{sec:method_overview}
A typical agent system processes user queries by interacting with a diverse set of tools and services within its environment to accomplish user tasks.
These tools may include code execution environments, email systems, web browsers, and file systems, among others.
The LLM in the agent serves as the planner, dynamically coordinating between these components to retrieve information, execute commands, and respond to user needs.
Given the complexity and autonomy of these systems, they often rely on external data sources, making them susceptible to various security threats.
The attacker exploits this reliance by strategically manipulating specific parts of the environment to inject malicious prompts.
These prompts are crafted to be embedded within external data sources, which the agent later retrieves and processes as part of its task execution.
Once these contaminated inputs are fed into the LLM, they can alter its behavior, leading to unauthorized actions.

Figure~\ref{fig:overview} illustrates the architecture and the workflow of our proposed framework, \sys.
\sys enhances the effectiveness of the indirect prompt attacks by systematically exploring adversarial prompts.
The process begins by applying the initial corpus of adversarial prompt templates to the agent across a set of injection tasks, which are combinations of different user tasks and attacker goals, to generate a pool of initial seeds.
These seeds then undergo an iterative fuzzing loop.
In this loop, a MCTS-based seed selector identifies a promising seed, balancing the dual objectives of exploitation and exploration.
Subsequently, a seed mutator randomly selects a mutation method to produce a new variant, which is then tested across the tasks.
This variant is subsequently tested across the injection tasks to evaluate its performance.
The evaluation involves scoring the new seed based on its success rate in executing attacks and its ability to compromise previously unaffected tasks.
Through this adaptive and iterative process, the framework continuously improve the attack, ensuring scalability and effectiveness across a wide range of agents and tasks.

\subsection{Corpus Collection}
\label{sec:corpus_collection}
To build a high-quality initial corpus, we collect adversarial prompt templates from a variety of sources, including human heuristics, online resources, existing prompt injection research~\cite{debenedetti2024agentdojo,liu2024formalizing}.
These templates are designed with placeholders to accommodate different variables, such as the specific LLM model in use, the user's task, and the attacker's goal, allowing for dynamic adaptation across different scenarios.
The corpus incorporates diverse attack strategies, including role-playing techniques where the model is coerced into adopting a specific persona, delimiter-based attacks that exploit structured inputs, and prompt obfuscation methods to bypass detection mechanisms.
By leveraging this diverse set of attack strategies, our framework ensures broad coverage of potential vulnerabilities, providing a strong foundation for the iterative fuzzing process to refine and optimize attack effectiveness.

\subsection{Mutation Design}
\label{sec:mutation_design}
Consistent with prior work \cite{yu2023gptfuzzer, yu2024promptfuzz}, we employ five mutation methods with prompt templates to prompt a helper LLM to generate new seeds based on existing seeds.
\textit{Shorten} compresses the seed for conciseness, \textit{Expand} adds additional contextual information, and \textit{Rephrase} introduces linguistic variety while preserving meaning.
\textit{Crossover} synthesizes elements from two parent seeds, and \textit{GenerateSimilar} prompts the creation of a stylistically similar seed with different content.
The mutations are randomly chosen for seed mutation at each iteration. 
We exclusively use basic mutation strategies without introducing extra heuristics, to maintain simplicity while encouraging diversity.
This approach ensures that the mutation process explores a broad range of variations without imposing additional constraints or biases on the generated seeds.
Furthermore, these basic mutation strategies require only moderately capable language models with smaller parameter sizes, such as Llama-3-8B and GPT-4o-mini. This allows for more efficient execution while still achieving diverse and meaningful mutations.

\subsection{Seed Scoring}
\label{sec:seed_scoring}
Our seed scoring strategy employs a hybrid evaluation mechanism that combines attack success rate (ASR) with coverage-guided assessment to identify and prioritize effective injection templates.
As detailed in Algorithm~\ref{alg:seed_scoring}, each seed undergoes performance evaluation across attack tasks, where the scorer monitors both the immediate success of attacks and the seed's contribution in broadening attack coverage across the overall task set. 
The final score is computed as a weighted sum of two components: the attack success rate, which is the ratio of successful attacks to total tasks, and a coverage bonus, which rewards seeds that uncover new successful attacks for previously failed ones.
This dual-metric approach ensures that seeds are valued for both their immediate effectiveness and their potential to explore new attacks.
Consequently, the framework maintains a balance between exploiting known successful patterns and exploring untapped attack patterns.
The coverage bonus term specifically incentivizes the discovery of injection patterns that work across diverse task contexts, promoting the development of more generalizable attack strategies.

\subsection{Seed Selection}
\label{sec:seed_selection}
Our framework utilizes a MCTS-based approach to intelligently navigate the space of injection templates by maintaining a tree structure that records mutation histories and relationships between seeds.
As shown in Algorithm~\ref{alg:seed_selection_select}, the selection mechanism utilizes the Upper Confidence Bound 1 (UCB1) algorithm \cite{Auer2002UCB1} to balance exploitation of high-scoring seeds with exploration of promising new variants.
For each node in the tree, the UCB score combines the node's empirical performance (exploitation term) with an exploration bonus that scales with the logarithm of total visits and inversely with the node's visit count.
This exploration term ensures that less-visited but potentially valuable branches of the mutation tree receive adequate attention. Given that the evaluation of each new seed is computationally expensive, we prioritize UCB1 over UCT to efficiently balance exploration and exploitation without requiring deep tree expansion.
Following each evaluation, Algorithm~\ref{alg:seed_selection_update} propagates visit counts up the ancestor chain, allowing the exploration bonus to naturally decay for well-explored mutation paths.
When selecting seeds for mutation, it selects the top-scoring one or two seeds based on the mutation strategies.
This MCTS-based selection strategy helps the framework efficiently identify and exploit promising mutation trajectories while maintaining sufficient diversity in the exploration process.
\section{Evaluation}
\label{sec:evaluation}
In this section, we comprehensively evaluate the effectiveness of \sys through the following analyses:
\begin{enumerate}[leftmargin=*,topsep=0pt]
    \setlength\itemsep{0em}
    \item We evaluate \sys on two estabilished agent benchmarks, AgentDojo~\cite{debenedetti2024agentdojo} (Section~\ref{sec:expr_agentdojo}), representing personal assistant agents, and VWA-adv~\cite{wu2024dissecting} (Section~\ref{sec:expr_vwa}), representing web-based agents, covering a variety of agent types and tasks.
    \item We evaluate the transferability of adversarial prompts generated by \sys across different LLMs and different tasks (Section~\ref{sec:expr_agentdojo}\&\ref{sec:expr_vwa}).
    \item We evaluate the effectiveness of adversarial prompts generated by \sys perform against various defense strategies deployed in the two benchmarks (Section~\ref{sec:expr_agentdojo}\&\ref{sec:expr_vwa}).
    \item We perform an ablation study to understand the contribution of key components of \sys (Section~\ref{sec:expr_ablation}).
    \item We examine the generated adversarial prompts in practical, real-world settings to demonstrate its applicability beyond controlled benchmark environments (Section~\ref{sec:expr_realworld}).
\end{enumerate}

Detailed versions of models used in our experiments are listed in Appendix~\ref{app:model_checkpoints}.

\subsection{Attack Personal Assistant Agents}
\label{sec:expr_agentdojo}
\paragraph{Experiment setup.}
In this section, we evaluate \sys using the AgentDojo framework~\cite{debenedetti2024agentdojo}, which is specifically designed for assessing indirect prompt injection attacks and defenses.
AgentDojo comprises several components: the environment, which defines an application area for an AI agent along with a set of available tools (such as a workspace environment with email, calendar, and cloud storage access); and the environment state, which tracks data for all applications the agent can interact with.
Certain parts of the environment state are specified as placeholders for potential indirect prompt injection attacks.
A user task is a natural language user query that the agent is expected to execute within the given environment (e.g., adding an event to a calendar), while an injection task outlines the attacker's objective (e.g., extracting the user’s credit card information). The collection of user tasks and injection tasks for an specific environment is referred to as a task suite.
AgentDojo provides formal evaluation criteria to assess the state of the environment, thereby measuring the success of both user and injection tasks.
In our context, a specific attack scenario or an adversarial task is defined as the combination of a user task and an injection task.
\sys interacts with AgentDojo by proposing adversarial prompts, which are then inserted into the placeholders in the environment for injection. The agent is subsequently run, and AgentDojo evaluates the success of the user and injection tasks. The success of the injection tasks serves as the attack success signal, providing feedback to \sys.

To evaluate the fuzzing performance and quality of the adversarial prompts generated by \sys, we randomly dividing the adversarial tasks within each suite of AgentDojo into two groups: a fuzzing set and a test set, with 142 and 173 tasks respectively.
We utilize GPT-4o-mini as the helper model to mutate the prompts in \sys.
We conduct the fuzzing experiment on the fuzzing set for the agent which utilizes the o3-mini model as the backbone due to its state-of-the-art reasoning capabilities.
We generate 3 mutated prompts in each iteration and complete a total of 10 fuzzing iterations.
Due to the large number of tasks, we randomly sample a quarter of user and injection tasks from each suite to evaluate each newly mutated seeds.
For the transferability experiment, we select the 5 seeds with the highest scores.
We evaluate the attack performance of the adversarial prompts, against o3-mini, GPT-4o, GPT-4o-mini, and Claude-3.5-Sonnet on the test set.
The success rate is computed on the union of the adversarial prompts.
According to AgentDojo, the Gemini and DeepSeek families and other open-source models do not fully support the tool call functionality  or are not as capable as the aforementioned LLMs.
We use the handcrafted adversarial prompts
proposed in AgentDojo as the baseline attack.
Furthermore, we assess the effectiveness of the generated adversarial prompts against defenses proposed in AgentDojo on the fuzzing set.
The defenses includes:
\textit{pi\_detector}~\cite{deberta-v3-base-prompt-injection-v2} utilizes a BERT classifier from ProtectAI to detect prompt injection;
\textit{repeat}~\cite{learning_prompt_sandwich_url} repeats the user instructions after each function call;
\textit{delimit}~\cite{hines2024defending} formats all tool outputs with special delimiters and incorporates system prompts to prioritize user instructions.
We exclude the \textit{tool\_filter}~\cite{simon2023dualllm} defense proposed in AgentDojo due to incompatibility with the o3-mini model.
We exclude other defenses due to several key reasons: they struggle to maintain utility, and face issues of high computational costs and adaptability.
For example, StruQ~\cite{chen2024struq} is demonstrated only on small open-source models, which lack the capability for agent tasks. Similarly, IsolateGPT~\cite{wu2025isolategpt} relies on a system-specific design that cannot be easily adapted to different agent architectures.

\paragraph{Fuzzing results.}
\begin{figure}[hptb]
    \centering
    \includegraphics[width=.9\linewidth]{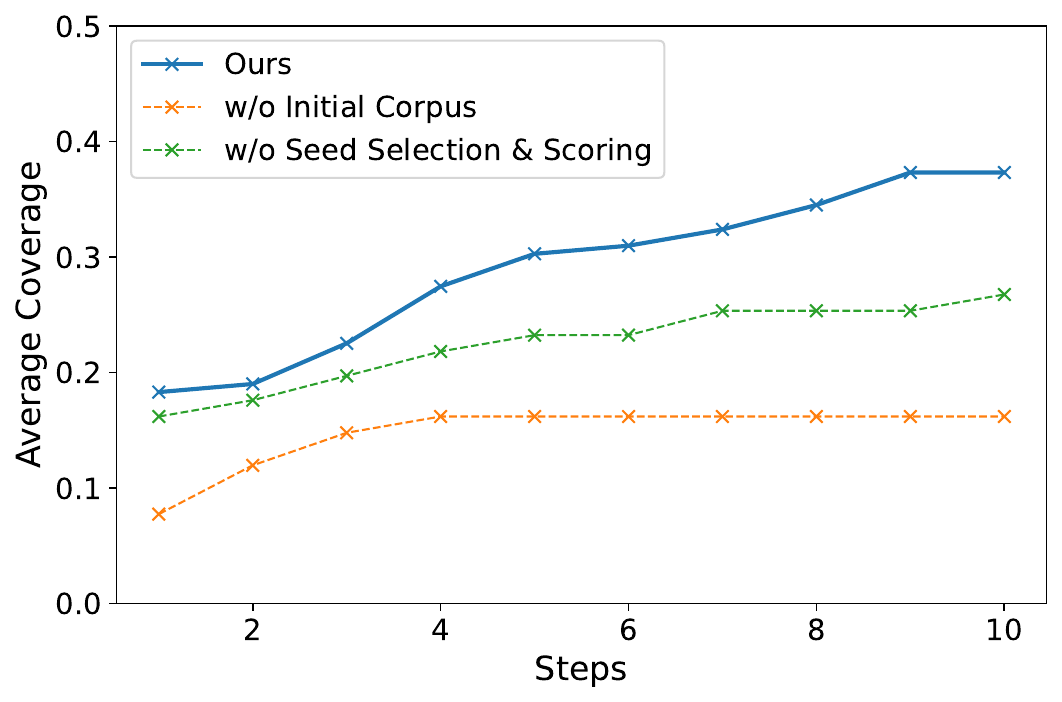}
    \caption{Coverage over fuzzing iteration steps achieved by \sys (the solid line) on AgentDojo with two ablation settings (the dashed lines): (1) without the high-quality initial corpus, (2) without the adaptive seed scoring strategy and the MCTS-based seed selection.}
    \label{fig:coverage_ablation}
\end{figure}
Figure~\ref{fig:coverage_ablation} presents the coverage progression over the course of the fuzzing iteration steps for \sys.
As shown, \sys continuously enhances the performance of the attack, resulting in higher coverage throughout the fuzzing process.
In terms of attack success rates, we compare \sys against the baseline handcrafted attacks in AgentDojo, which achieve a success rate of 38\%.
Our initial high-quality corpus demonstrates a 63\% success rate, showcasing its ability to surpass baseline prompts.
As fuzzing iterations progress and the adversarial prompts are further refined, \sys achieves a 71\% success rate—a significant improvement over both the baseline and the initial corpus.
These findings underscore the efficacy of adaptive fuzzing for uncovering injection vulnerabilities in blackbox agents and highlight the effectiveness of targeted search strategies in maximizing the attack performance.

\paragraph{Transferability.}
\begin{table*}[htbp]
  \centering
  \caption{The transfer attack success rate of selected adversarial prompts generated by \sys compared with the baseline attacks proposed by AgentDojo~\cite{debenedetti2024agentdojo} and VWA-adv~\cite{wu2024dissecting}, against the agents using different backbone LLMs. We run fuzzing against o3-mini on AgentDojo, GPT-4o on VWA-adv.
  }

  \begin{threeparttable}
    \resizebox{.85\linewidth}{!} {

    \begin{tabular}{cccccccc}
    \toprule
    \multirow{2}[1]{*}{Benchmark} & \multirow{2}[1]{*}{Task set} & \multirow{2}[1]{*}{Attack} & \multicolumn{5}{c}{Model} \\ \cline{4-8}
    & & & o3-mini & GPT-4o & GPT-4o-mini & Claude-3.5-Sonnet & Gemini-2-flash-exp \\ \hline
    \multirow{4}[1]{*}{AgentDojo} & \multirow{2}[1]{*}{Fuzzing} & handcrafted &  0.38   &  0.22     &   0.28   &   \textbf{0.12}    &   - \\
         &  & \sys  &   \textbf{0.71}   &   0.22    &   \textbf{0.49}    & 0.03 & -\\ \cline{2-8}
    & \multirow{2}[1]{*}{Test} & handcrafted &    0.34    & \textbf{0.25}  &  0.28     &  \textbf{0.08}     &  - \\
          & & \sys  &   \textbf{0.65}   &   0.19    &  \textbf{0.43}     & 0.04      & - \\ \hline
    \multirow{4}[1]{*}{VWA-adv} & \multirow{2}[1]{*}{Fuzzing} & handcrafted & - &       0.36 &    0.08  &  \textbf{0.47} & 0.49  \\
         &  & \sys  &    -  &   \textbf{0.60}   &   \textbf{0.47}    & 0.31      & \textbf{0.67}  \\ \cline{2-8}
    & \multirow{2}[1]{*}{Test} & handcrafted &    -  &  0.44     &   0.29    &  \textbf{0.51}      & 0.50 \\
          & & \sys  &  -    &   \textbf{0.59}    &   \textbf{0.54}    &   0.42    & \textbf{0.67} \\
    \bottomrule
    \end{tabular}
      }
    \begin{tablenotes}
   \small \item[1]Gemini family doesn't fully support the tool calls in AgentDojo. Early version of o3-mini doesn't fully support VWA-adv framework.
    \end{tablenotes}
    \end{threeparttable}

      \label{tab:agentdojo_vwa_transfer}%
      \vspace{-2mm}
\end{table*}%

As shown in Table~\ref{tab:agentdojo_vwa_transfer}, the success rate results in the first column for o3-mini on the test task set indicate that the generated adversarial prompts transfer effectively across different tasks, even with varying user tasks and injection goals, significantly outperforming the baseline attack—nearly doubling its performance.
Comparing performance across rows, we observe that the adversarial prompts transfer well to GPT-4o-mini but perform relatively worse on GPT-4o and Claude-3.5-Sonnet. 
Furthermore, both the baseline and \sys's prompts are ineffective against Claude-3.5-Sonnet, as it demonstrates strong robustness in defending against complex adversarial prompts.

\paragraph{Against defenses.}
\begin{table}[htbp]
  \centering
  \caption{The attack success rate of selected adversarial prompts generated by \sys on fuzzing task set and o3-mini against four defenses proposed by AgentDojo.}
  \resizebox{\linewidth}{!} {
    \begin{tabular}{cccccc}
    \toprule
    \multirow{2}[1]{*}{Attack} & \multirow{2}[1]{*}{No Defense} & \multicolumn{3}{c}{Defenses} \\ \cmidrule(lr){3-5}
    & & pi\_detector & repeat & delimit \\ \midrule
    baseline & 0.38 & 0.13 & \textbf{0.21} & 0.36 \\
    \sys & \textbf{0.71} & \textbf{0.25}  & 0.12 & \textbf{0.49} \\
    \bottomrule
    \end{tabular}
  }
  \label{tab:agentdojo_defense}%
\end{table}%
The results in Table~\ref{tab:agentdojo_defense} demonstrate the effectiveness of \sys against defenses compared to the baseline.
Checking along the columns, \sys consistently outperforms the baseline, particularly against \textit{pi\_detector} and \textit{delimit}, indicating that adversarial prompts generated by \sys are more resilient to these defenses.
Examining the results along the rows, both the baseline and \sys experience significant drops in success rates when defenses are applied.
However, the attacks still maintain high success rates, highlighting the insufficiency of these defenses.
Additionally, according to the results, \textit{delimit} is less effective than \textit{pi\_detector} and \textit{repeat}, as both \sys and baseline achieve higher success rate against \textit{delimit} among the defenses.

\subsection{Attacking Web Agents}
\label{sec:expr_vwa}
\paragraph{Experiment setup.}
In this section, we further evaluate \sys on VWA-adv~\cite{wu2024dissecting}.
VWA-adv is a set of realistic adversarial tasks based on VisualWebArena~\cite{koh2024visualwebarena}, which serves as a benchmark for evaluating web agents on a set of diverse and complex web-based visual tasks with multi-modal input.
Each task in VWA-adv consists of an original task in VisualWebArena and a trigger image or trigger text, which serves as the injection point, along with a targeted adversarial goal as the attacker’s objective.
In VWA-adv, attacker goals fall into two categories: illusioning, which misleads agents about object attributes (e.g., changing an object's color), and goal misdirection, which alters the agent's intended action (e.g., adding an item to the cart).
We focus on the tasks with text trigger.
Similar to Section~\ref{sec:expr_agentdojo}, we feed the adversarial prompts from \sys to the evaluation framework in VWA-adv, which then returns whether the adversarial task succeeds or not as feedback to \sys.

Similarly, we randomly divide the tasks in VWA-adv into a fuzzing set (99 tasks) and a test set (100 tasks) to evaluate the fuzzing performance and quality of the generated adversarial prompts, respectively.
We utilize GPT-4o-mini as the helper model to mutate the prompts in \sys.
We run the fuzzing experiment against the agents using GPT-4o on the fuzzing set.
We generate 10 mutated prompts per iteration and conduct 10 iterations in total.
We use the handcrafted adversarial prompts proposed in VWA-adv as the baseline.
We select 5 seeds with the highest scores to conduct the transferability experiment.
We evaluate the attack performance of the adversarial prompts against GPT-4o, GPT-4o-mini, Claude-3.5-Sonnet, and Gemini-2-flash-exp on the test set.
We further assess the effectiveness against basline defenses proposed in VWA-adv on the fuzzing set.
There are three defenses: \textit{safety}~\cite{hines2024defending} utilizes the data delimiter and system prompts to prioritize user instructions; \textit{paraphrase}~\cite{jain2023baseline} paraphrases untrusted text to neutralize malicious intent; \textit{combined} integrates both strategies.
While VWA-adv includes one more defense which checks consistency between image and text content, we exclude it since it would substantially increase API calls, making it impractical for real-world use.

\vspace{-3mm}
\paragraph{Fuzzing results.}
Figure~\ref{fig:vwa_coverage} shows \sys's coverage progression during fuzzing. \sys steadily improves attack performance, achieving higher coverage.
Compared to baseline attacks in VWA-adv with a success rate of {36\%}, our high-quality initial corpus starts at {54\%} and surpasses the baseline.
With iterative refinement, \sys reaches {70\%}, nearly doubling the baseline’s success rate and significantly outperforming both.
These results demonstrate \sys's effectiveness in exposing injection vulnerabilities and optimizing attack performance.

\vspace{-1mm}
\paragraph{Transferability.}
The lower half of Table~\ref{tab:agentdojo_vwa_transfer} presents the attack success rates of adversarial prompts from \sys compared to the VWA-adv baseline.
The results demonstrate that \sys significantly outperforms the baseline, achieving an absolute success rate improvement of 15\% to 40\% across different models and tasks except Claude-3.5-Sonnet.
This highlights the high quality and effectiveness of the adversarial prompts, as well as their strong transferability.
Consistent with Section~\ref{sec:expr_agentdojo}, adversarial prompts optimized for GPT do not transfer well to Claude, whereas baseline attacks from VWA-adv achieve higher success rates on Claude compared to other models.
Upon manual inspection, we suspect that Claude is more vulnerable to simpler adversarial prompts, differing from the GPT family.
Furthermore, the findings reinforce the conclusion from VWA-adv that prompt injection is an effective attack capable of overriding the influence of visual input on the model.
It is worth noting that \sys achieve approximately 50\% and 60\% on GPT-4o-mini and GPT-4o, respectively, suggesting that the instruction hierarchy~\cite{wallace2024instruction} defense mechanism is not sufficiently effective.

\paragraph{Against defenses.}
\begin{table}[thbp]
  \centering
  \caption{The attack success rate of selected adversarial prompts generated by \sys on fuzzing task set and GPT-4o against three defenses proposed by VWA-adv.}
  \resizebox{\linewidth}{!} {
    \begin{tabular}{ccccc}
    \toprule
    \multirow{2}[1]{*}{Attack} & \multirow{2}[1]{*}{No Defense} & \multicolumn{3}{c}{Defenses} \\ \cmidrule(lr){3-5}
    & & safety & paraphrase & combined \\ \midrule
    baseline & 0.36 & \textbf{0.34} & 0.27 & \textbf{0.30} \\
    \sys & \textbf{0.60} & 0.29 & \textbf{0.33} & 0.27 \\
    \bottomrule
    \end{tabular}
  }
  \label{tab:vwa_defense}%
\end{table}%
The evaluation results in Table~\ref{tab:vwa_defense} of \sys highlight a substantial improvement in attack success when no defense mechanisms are applied, achieving a 60\% success rate compared to the baseline’s 36\%.
However, when defenses are introduced, \sys's performance declines and converges with the baseline.
This degradation is likely due to the complexity of the attack prompts, which, while effective in an unprotected setting, struggle against the defenses due to the limited context in VWA-adv.
Notably, we observe that the combined defense does not further reduce the attack success rate compared to individual defenses. 
This suggests that certain attack prompts are inherently more robust and can bypass multiple defenses simultaneously, indicating potential weaknesses in the current defense mechanisms.

\subsection{Ablation Study}
\label{sec:expr_ablation}
We perform an ablation study on AgentDojo to isolate the impact of each of the three core components of \sys: the initial corpus of adversarial prompt templates, the adaptive seed scoring strategy, and the MCTS-based seed selection.
Specifically, we (1) replace our initial corpus with the handcrafted baseline prompts from AgentDojo, (2) substitute the adaptive seed scoring and MCTS-based seed selection to uniform random seed selection.
As shown in Figure~\ref{fig:coverage_ablation}, \sys significantly outperforms the ablated versions.
Notably, when the initial corpus is replaced by the baseline prompts, the overall success rate plateaus after approximately four iterations, demonstrating both reduced performance and limited potential compared to our curated initial corpus. 
Furthermore, without adaptive seed scoring or the MCTS-based seed selection, the fuzzing process shows markedly slower improvement, as it fails to identify and prioritize high-potential seeds.
These findings underscore the critical role of all three components in driving \sys's continuous enhancement and superior attack success.

\subsection{Real-world Case Study}
\label{sec:expr_realworld}
Figure~\ref{fig:realworldcase} shows the workflow of the indirect prompt injection for a web agent in the real world.
In this experiment, we deploy a shopping website provided by WebArena~\cite{zhou2023webarena} and use the default agent implementation in WebArena. The shopping website in WebArena is based on a famous open source e-commerce project magento2~\cite{magento2} which has many real-world deployment instances.
Due to ethical considerations, we use a local copy in this experiment.
As shown in the figure, the user task is to find a screen protector and list out reviewers who mention good fingerprint resistant.
This user task involves first searching for the product then reading the customer reviews of the target product with over ten step operations in total.
The attacker left a review with malicious prompts, which can lead to undesired actions.
Here we use the generated adversarial prompts in Section \ref{sec:expr_vwa} and inject it to the customer reviews by a regular user account like a normal customer.
In the figure, we use a fake URL of GitHub as an example, which is a commonly used pattern of phishing sites, and the results show that our attack method can lead the agent to visit arbitrary URLs including visiting phishing sites, downloading malicious files, sending out private information.
This case study proves the results in the previous experiments can be transfered to a more real-world scenario.

\section{Conclusion}
\label{sec:conclusion}
We introduce \sys, a novel fuzzing framework designed to systematically conduct indirect prompt injection attacks against blackbox agents with various architectures and tasks.
By combining high-quality prompt templates, adaptive seed scoring, and a MCTS-based seed selection algorithm, \sys overcomes challenges posed by the black-box nature, architectural complexity, and wide-ranging functionalities of real-world agents.
Our empirical results demonstrate that \sys not only achieves high attack success rates on established benchmarks and real-world agents but also exhibits strong transferability across unseen tasks and underlying LLMs.
By automating the generation and optimization of adversarial prompts, \sys highlights critical limitations in existing agent defenses, underlining the urgent need for more robust security measures.
We believe \sys will serve as a useful foundation for advancing both the understanding of agent-based threats and the development of next-generation security solutions in this rapidly evolving domain.

\section*{Impact Statement}
This work provides a significant advancement in uncovering the security vulnerabilities of LLM-based agent systems by exposing how indirect prompt injection attacks can be launched even under blackbox constraints. Although our fuzzing framework is primarily an offensive testing tool, its results offer vital insights for agent developers and security researchers, guiding the development of more robust defense mechanisms and secure system designs. By revealing weaknesses early, we help stakeholders protect against malicious manipulations while enabling the legitimate and safe use of agent systems in real-world settings. Nonetheless, no single testing or defense approach is infallible; ongoing research and proactive updates remain essential to address evolving threats in this dynamic landscape.


\bibliography{ref}
\bibliographystyle{icml2025}

\appendix
\onecolumn
\section{Detailed Model Checkpoints}
\label{app:model_checkpoints}
The models used in our evaluation use the following checkpoints:
o3-mini (o3-mini-2024-12-17), GPT-4o-mini (gpt-4o-mini-2024-07-18), GPT-4o (gpt-4o-2024-08-06), Claude-3.5-Sonnet (claude-3-5-sonnet-20241022), Gemini-2-flash-exp (gemini-2.0-flash-exp).

\section{Additional Evaluation Results}
\subsection{Evaluation on Extra Models}

While our primary focus has been on commercial black-box LLMs such as GPT, Claude, and Gemini, we also evaluate \sys on open-source models. 
These models often lag behind in long-context understanding, advanced tool usage, reasoning, and planning, which are essential in the challenging agent scenarios.
We assess models that support tool calling using the AgentDojo benchmark and report their utility scores (i.e., success rates on benign tasks): {Llama3.3-70B-Instruct}~\cite{llama33} (42\%), {Qwen2.5-72B-Instruct}~\cite{qwen2.5} (54\%), {QwQ-32B}~\cite{qwq32b} (74\%), and, for comparison, {o3-mini} (79\%). Based on these results, we conduct further experiments using {QwQ-32B}.
Additionally, the {o3-mini} checkpoint used in the main text corresponds to an experimental version. We therefore also evaluate the latest available checkpoint, {o3-mini-2025-01-31}.
Results indicate that \sys achieves a success rate of 72\% on the fuzzing set and 74\% on the test set, compared to the baseline handcrafted attack, which achieves 50\% and 53\%, respectively, as shown in Table~\ref{tab:QwQ_and_more_baselines}. These findings underscore \sys's effectiveness, even when applied to strong open-source models.

\subsection{Comparison with Additional Baselines}

To enhance our baseline comparisons, we included two additional prompt injection baselines from OpenPromptInjection~\cite{liu2024formalizing} and InjecAgent~\cite{zhan2024injecagent}.
Table~\ref{tab:QwQ_and_more_baselines} reports the attack success rates on the Fuzzing and Test sets across two models, QwQ-32B and o3-mini.
These results confirm that \sys outperforms state-of-the-art baselines, especially in discovering and exploiting indirect prompt injection vulnerabilities.

\begin{table}[H]
\centering
\captionsetup{width=.8\linewidth}
\caption{Attack success rate (ASR) comparison on AgentDojo with \sys and three baseline attacks. The two ASRs in each cell represent performance on the Fuzzing task set and Test task set, respectively (i.e., \textit{Fuzzing} / \textit{Test}).}
\label{tab:QwQ_and_more_baselines}
\begin{tabular}{ccc}
\toprule
\multirow{2}[1]{*}{Attack} & \multicolumn{2}{c}{Model} \\ \cline{2-3}
& o3-mini-2025-01-31 & QwQ-32B \\
\midrule
   \sys           & \textbf{0.73} / \textbf{0.76} & \textbf{0.72} / \textbf{0.74}\\
   AgentDojo Baseline    & 0.47 / 0.49 & 0.45 / 0.47 \\
   OpenPromptInjection & 0.38 / 0.39 & 0.20 / 0.20 \\
   InjecAgent            & 0.15 / 0.11 & 0.14 / 0.12 \\
\bottomrule
\end{tabular}
\end{table}

\subsection{Breakdown on Attack Scenarios}
The two benchmarks, AgentDojo and VWA-adv, are designed to evaluate performance across diverse scenarios. We perform additional analysis about the detailed results on different scenarios to provide a comprehensive view of \sys's effectiveness.

AgentDojo consists of various agent tasks grouped into four suites -- Slack, Workspace, Travel, and Banking. As shown in Table~\ref{tab:scenario}, across all these scenarios, \sys consistently achieves a higher success rate compared to the baseline attacks in AgentDojo, demonstrating its robustness and adaptability in different operational environments.

On VWA-adv benchmark, we evaluate performance across two types of adversarial goals: illusioning, which makes it appear to the agent that it is in a different state (e.g., different objects, colors), and goal misdirection, which makes the agent pursue a targeted different goal than the original user goal (e.g., leave a comment). Our results in Table~\ref{tab:scenario} indicate that \sys outperforms baseline from the benchmark in both attack goals, confirming its capability to exploit diverse indirect prompt injection vulnerabilities and attack goals effectively, even in challenging goal misdirection tasks.

\begin{table}[H]
\centering
\caption{Attack success rates across different scenarios (task suites for AgentDojo, attack goals for VWA-adv) achieved by \sys and the baseline attack from the benchmark. The two ASRs in each cell represent performance on the Fuzzing task set and Test task set, respectively (i.e., \textit{Fuzzing} / \textit{Test}).}
\label{tab:scenario}
\begin{tabular}{ccccc}
\toprule
Benchmark & Model & Scenario & \sys & Benchmark Baseline \\
\midrule
\multirow{8}[1]{*}{AgentDojo} & \multirow{4}[1]{*}{o3-mini}  & Slack     & 0.81 / 0.97 & 0.64 / 0.70 \\
         && Workspace & 0.63 / 0.60 & 0.20 / 0.22 \\
         && Travel    & 0.71 / 0.83 & 0.55 / 0.50 \\
         && Banking   & 0.49 / 0.38 & 0.25 / 0.23 \\
\cmidrule(lr){2-5}
&\multirow{4}[1]{*}{QwQ-32B}  & Slack     & 1.00 / 0.97 & 0.85 / 0.88 \\
         && Workspace & 0.33 / 0.42 & 0.05 / 0.10 \\
         && Travel    & 0.80 / 0.80 & 0.60 / 0.65 \\
         && Banking   & 0.60 / 0.65 & 0.23 / 0.23 \\
\midrule

\multirow{2}[1]{*}{VWA-adv} & \multirow{2}[1]{*}{gpt-4o} & Illusioning       & 0.82 / 0.76 & 0.51 / 0.62 \\
&& Goal misdirection & 0.58 / 0.42 & 0.00 / 0.20 \\
\bottomrule
\end{tabular}
\end{table}

\subsection{Coverage Curve}
\begin{figure}[H]
    \centering
    \includegraphics[width=.5\linewidth]{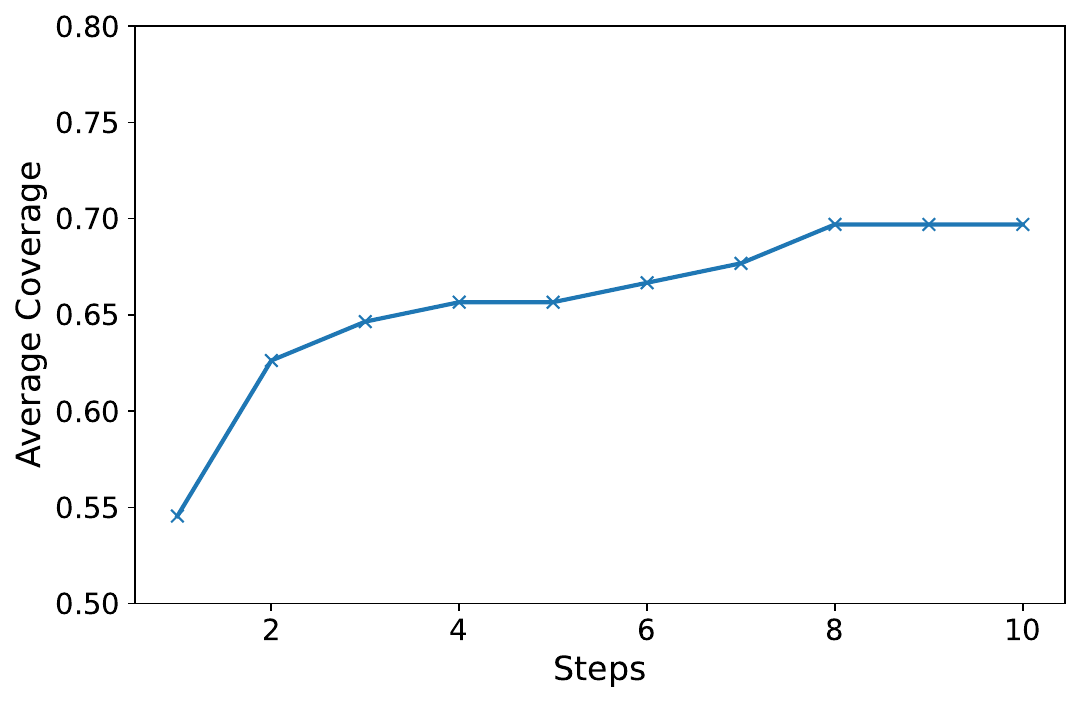}
    \caption{Coverage over fuzzing iterations achieved by \sys on VWA-adv.}
    \label{fig:vwa_coverage}
\end{figure}

The coverage of tasks over fuzzing iterations achieved by \sys on VWA-adv benchmark is shown in Figure~\ref{fig:vwa_coverage}.

\section{Algorithms}
The algorithms for our seed scoring and selection are shown in \cref{alg:seed_scoring,alg:seed_selection_select,alg:seed_selection_update}.
\begin{algorithm}[tbph]
\caption{Success rate and coverage-guided seed scoring
}
\label{alg:seed_scoring}
\begin{algorithmic}[1]
\REQUIRE Seed to be evaluated $\mathsf{seed}$, coverage factor $C$
\ENSURE Final score for the seed and suite results
\STATE /* Initialize */
\STATE $\mathsf{total\_success} \gets 0$
\STATE $\mathsf{num\_questions} \gets 0$
\STATE $\mathsf{coverage\_bonus} \gets 0$

\FORALL{$\mathsf{task\_suite}$ in $\mathsf{sampled\_tasks}$}
    \STATE /* Evaluate user and injection task combinations using $\mathsf{seed}$. */
    \STATE /* Compute attack success rate for the suite. */
    \IF{injection successful}
        \STATE Increment $\mathsf{total\_success}$.
    \ENDIF
    \STATE Increment $\mathsf{num\_questions}$.
    \STATE /* Identify newly successful task combinations not covered before and: */
    \IF{injection successful}
        \STATE /* Mark combination as covered. */
        \STATE Increment $\mathsf{coverage\_bonus}$.
    \ENDIF
\ENDFOR

\STATE /* Calculate Final Score including attack success rate and coverage bonus. */
\STATE $ASR \gets \frac{\mathsf{total\_succcess}}{\mathsf{num\_questions}}$.
\STATE $\mathsf{seed\_score} \gets \mathsf{ASR} + C \cdot \frac{\mathsf{coverage\_bonus}}{\mathsf{num\_questions}}$.


\STATE \textbf{Return} $\mathsf{seed\_score}$

\end{algorithmic}
\end{algorithm}

\begin{algorithm}[tbph]
\caption{MCTS-based seed selection: Update}\label{alg:seed_selection_update}
\begin{algorithmic}[1]
\REQUIRE Set of nodes $N$, new node $\mathsf{node}$ with information of parent node(s) $\mathsf{node}.parents$ and the score $\mathsf{node}.score$.
\ENSURE Updated set of nodes $N$
\STATE /* Update all the ancestors of the node */
\STATE $\mathsf{ancestors} \gets \mathsf{node}.parents$
\FOR{ancestor $p \gets \mathsf{ancestors}.pop()$}
    \STATE $p.visits \gets p.visits + 1$
    \STATE $\mathsf{ancestors} \gets \mathsf{ancestors} \cup p.parents$
\ENDFOR
\STATE /* Update the set of nodes */
\STATE $N \gets N \cup \{\mathsf{node}\}$

\STATE \textbf{Return} $N$
\end{algorithmic}
\end{algorithm}
\begin{algorithm}[tbph]
\caption{MCTS-based seed selection: Select}\label{alg:seed_selection_select}
\begin{algorithmic}[1]
\REQUIRE Set of nodes $N$, exploration factor $C$, number of nodes to select $n$
\ENSURE Selected node(s) $S$
\STATE $\mathsf{total\_visits} \gets \sum_{\mathsf{node} \in N} \mathsf{node}.visits$
\STATE $\mathsf{UCB}(\mathsf{node}) \gets \mathsf{node}.score + C \cdot \sqrt{\frac{\log(\mathsf{total\_visits} + 1)}{\mathsf{node}.visits + \epsilon}}$
\IF{$n = 1$}
    \STATE Select $S \gets \arg\max_{\mathsf{node} \in N} \mathsf{UCB}(\mathsf{node})$
\ELSIF{$n = 2$}
    \STATE Sort $N$ by $\mathsf{UCB}(\mathsf{node})$ in descending order
    \STATE Select $S \gets$ top 2 nodes in $N$
\ENDIF

\STATE \textbf{Return} $S$
\end{algorithmic}
\end{algorithm}

\end{document}